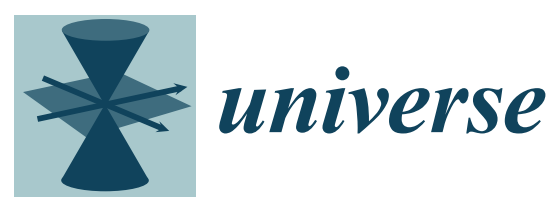

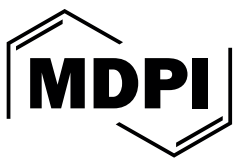

Article

# Simulation of the Spin Evolution of Some Selected Exoplanets and Inferences on Their Climate

**Salvatore Camposeo [1,2,3,*], Francesco De Paolis [4,5,6], Vincenzo Orofino [4,5,6], Francesco Strafella [4,6] and Leonardo Di Venere [3]**

1 Dipartimento di Fisica e Astronomia "Galileo Galilei", Università di Padova, 35131 Padova, Italy
2 Dipartimento Interuniversitario di Fisica "M.Merlin", Università di Bari, 70126 Bari, Italy
3 Istituto Nazionale di Fisica Nucleare (Sezione di Bari), 70126 Bari, Italy; leonardo.divenere@ba.infn.it
4 Dipartimento di Matematica e Fisica "E. De Giorgi", Università del Salento, 73100 Lecce, Italy; francesco.depaolis@le.infn.it (F.D.P.); vincenzo.orofino@le.infn.it (V.O.); francesco.strafella@unisalento.it (F.S.)
5 Istituto Nazionale di Fisica Nucleare (Sezione di Lecce), 73100 Lecce, Italy
6 Istituto Nazionale di Astrofisica (Sezione di Lecce), 73100 Lecce, Italy
* Correspondence: salvatore.camposeo@ba.infn.it or salvatore.camposeo@phd.unipd.it

**Abstract**

In this work, using the simulator *VPLanet*, we analyze the spin evolution of some selected exoplanets due to the tidal interaction with their host star. For a rocky planet, two spin "conditions" are possible, the "trapped" rotation and the "fast" rotation, referring to the cases of achieved and non-achieved tidal trapping, respectively. We focus on planets whose spin condition is not obvious, because no study is needed for planets which are undoubtedly fast rotators or undoubtedly trapped rotators; moreover, we consider only exoplanets that are interesting from an astrobiological perspective. The current spin conditions of the considered planets are hypothesized, taking into account the age of the host star. Inferences regarding planetary climate and habitability—which is defined by the possibility of stably sustaining the liquid water on the surface—are also discussed. Results of this work show that Kepler-62f, Kepler-1126c, and Kepler-1544b are expected to be fast rotators regardless of the orbital eccentricity; the spin condition of Kepler-186f, Kepler-62e, and Kepler-442b cannot be determined without constraints on the eccentricity, which are currently unavailable; Kepler-440b is expected to be tidally trapped.

## 1. Introduction

The existence of over 6000 exoplanets has been confirmed until now (https://exoplanetarchive.ipac.caltech.edu/, latest access on 27 February 2026). Among these, some are rocky and placed in their theoretical habitable zone, which makes them astrobiologically interesting. From a biological perspective, habitability is defined by the possibility of stably sustaining the liquid water on the surface [1,2].

In this work, we focus on a specific subgroup: that of astrobiologically interesting planets whose spin conditions are uncertain, meaning that their orbital period is greater than that of Mercury but smaller than that of the Earth.

In fact, Mercury is a planet whose spin was affected early in its history by its interaction with the host star [3], while the Earth is a planet whose spin condition has not been significantly affected by its host star over billions of years [4].

We expect planets having smaller orbital periods than Mercury and greater orbital periods than Earth to be tidally trapped and fast rotators, respectively. The specific age of the planet can affect its spin condition: for example, if the planet is very young, it has a low probability of being tidally trapped since not enough time has elapsed in order to reach any resonance condition. However, as will be shown in a later section, the time $\tau$ needed to achieve any resonance is so short for Mercury and so long for the Earth—compared to stellar evolution timescales—that the choice of these two limits is appropriate. In our solar system, the only planet having an intermediate orbit between that of Mercury and that of the Earth is Venus, but it cannot help us constrain further the range of $w$ values which we associate with an uncertain spin condition. In fact, the spin evolution of Venus has been very peculiar and has led the planet to a retrograde rotation, resulting from a combination of a giant impact and strong thermal tides [5–7]. The condition of Venus and also the ensemble of causes leading to it are expected to rarely occur, so we will ignore the case of Venus [5–7].

Among the group of astrobiologically interesting exoplanets, only seven have an orbital period between those of Mercury and Earth. We have focused on these seven objects and have tried to solve the uncertainty about their spin condition. In particular, with the recent code *VPLanet v2.5.36* [8,9], we have computed the time $\tau$ needed to achieve tidal trapping for each of these planets, and then we have compared these values with the age of the host stars.

The spin condition of a planet can be described by two macro classes [8,9]: "tidally trapped" and "fast rotator" (i.e., not trapped). The first class is itself divided into two subclasses: "1:1 resonance" (also known as "perfect resonance") and "3:2 resonance" (also known as "imperfect resonance"). The first subclass consists of planets whose surfaces "see" the host star as a fixed source in the sky. The second subclass consists of slow rotators—which are, in any case, true rotators—and they "see" the host star "slowly moving" in the sky. Once the current spin condition of a planet has been estimated (if possible), hypotheses about its climate can be made. This has been done for all seven of the selected exoplanets: Kepler-62f, Kepler-1126c, Kepler-1544b, Kepler-186f, Kepler-62e, Kepler-442b, and Kepler-440b.

The climate of a planet is certainly affected by its spin–orbit condition [1,2]. In fact, a planet with a spin–orbit frequency ratio of 1:1 has a "bimodal" climate, since the planet presents a permanently illuminated hemisphere and a complementary dark one. Also, a planet having a spin–orbit frequency ratio of 3:2 has a peculiar climatological behavior, which is different from that of a fast rotator (like Earth), as we will show.

Resonances 1:1 and 3:2 are the most likely conditions for a "tidally trapped" planet [5,10]. Two other resonances having a non-negligible (but very low) probability of being achieved are the 5:2 and the 2:1 ones, but they are expected to be equivalent to the 3:2 condition from a climatic point of view, because both involve a very slow rotation [1,2]. Consequently, only 1:1 and 3:2 resonances are investigated in this work. Let us observe that a planet experiencing an imperfect resonance completes three rotations every two orbits. The plan of the paper is as follows:

- In Section 2.1, we describe the data selection criteria;
- In Section 2.2, we briefly present the *VPLanet* code for the simulations of the spin evolution;
- In Section 2.3, we explain the approach of climatological simulations;
- In Section 2.4, we show the input data for both *VPLanet* simulations and climatological simulations;
- In Section 3, we show our numerical results; then, in Section 4, we briefly comment on and summarize all the simulation outcomes.

## 2. Methods

### 2.1. Data Selection Criteria

Let us define the "tidal acceleration parameter" as $w = \frac{M_{star}}{a^3}$: the tidal acceleration experienced by the planet is proportional to $w$, where $M_{star}$ is the mass of the host star and $a$ is the semi-major axis of the planet's orbit [5]. We are assuming that the planetary mass is negligible compared to that of its host star. For the sake of convenience, we will implicitly express $w$ in units of $w_{Earth} = 1\frac{M_\odot}{AU^3}$, where $M_\odot$ is the solar mass, and AU is the astronomical unit of distance. Note that, due to the third law of Kepler, one has $w \propto T_{orbital}^{-2}$, where $T_{orbital}$ is the orbital period.

Let us note that, if a massive satellite is present, it also induces tidal acceleration on the planet. However, in this work, we will always assume the absence of moons, because the presence of massive satellites having more than 1% of planetary mass (like our Moon) is expected to be very rare around rocky planets [11]. Furthermore, this reasonable assumption allows us to avoid all computational complications given by the presence of a massive satellite.

In this work, we selected a subpopulation of exoplanets within the general one (which is found at https://exoplanetarchive.ipac.caltech.edu/, latest access on 27 February 2026) on the basis of the following requirements.

As a first step, we have considered all confirmed transiting planets, whose radii and orbital periods are measured, orbiting a single main sequence star. The method of transit photometry—the one exploited by the famous *Kepler* telescope—is the best for measuring both the planetary radius and the orbital period with high precision, which are fundamental parameters in our study [12]. Planets orbiting evolved stars—i.e., non-main sequence stars—have been excluded because their study would require a further hypothesis on the star radius (whereas, for main sequence stars, the stellar radius is relatively small and cannot affect the evolution of planetary rotation [5]). Also, planets belonging to multiple star systems have been discarded, because the investigation of their spin evolution and climate would be extremely difficult.

As a second step, we have excluded all planets having $R > 2R_\oplus$, which are expected to be non-rocky worlds [13,14]. We omitted non-rocky planets because they have very deep atmospheres, so we would not be able to assess their input parameters (i.e., the second degree Love number $k_2$ and the tidal quality factor Q, which are defined in Appendix A) [5].

As a third step, all planets whose host star age, luminosity, or mass are unknown have also been excluded. Note that the stellar age is expected to be equal to the age of hosted planets, since protoplanetary disks around young stars are expected to dissolve very quickly [6,15].

As a fourth step, we have selected planets with $1.0\, w_\oplus < w < 17.4\, w_\oplus$, where these two edge values are the tidal acceleration parameters of Earth and Mercury, respectively.

We know that:

- Mercury has reached its stable resonance frequency ratio (3:2) in the past; many simulations (including ours, as we will see below) show that it has almost certainly been trapped in its current condition in less than 3 Gyr ($1\text{ Gyr} = 10^9$ years) after its formation [3];
- Earth has not experienced—for the 4.5 Gyr since its formation—a slowdown induced by the Sun; however, a spin slowdown of ~16 h has been estimated, but it has actually been induced by the Moon [4].

Given these considerations, we have decided to take $w_{Earth}$ and $w_{Mercury}$ as lower and upper limits for $w$, respectively. Smaller or higher values of $w$ (out of the considered range) would give obvious results, so investigation would be unnecessary: planets with $w < w_{Earth}$ are not trapped, while planets with $w > w_{Mercury}$ are surely trapped.

A fifth and final filtering is given by requiring that the planet lies in the classic "habitable zone", which is defined by an inner bound where the planet has a Venus-like insolation and an outer bound where the planet has a Mars-like insolation [16]. Insolation $F$ is the stellar flux arriving on the planet before absorption, scattering, and reflection and is usually expressed in units of Earth insolation $F_{\oplus} = 1361$ W/m$^2$ [6].

In the cases of Venus and Mars, the insolation values are 1.91 $F_{\oplus}$ and 0.43 $F_{\oplus}$, respectively [6]. However, we can extend the range of "acceptable" insolation values and consider an "optimistic" outer bound corresponding to 0.32 $F_{\oplus}$ [16]. In theory, a planet having a very large mass, a very strong magnetic field, and a very low *Albedo* could be habitable even with an insolation in the range 0.32 $F_{\oplus}$–0.43 $F_{\oplus}$ [16]. Hence, we are taking an insolation range between 0.32 $F_{\oplus}$ and 1.91 $F_{\oplus}$ as filtering criterion.

It is worth noting that a flux between 1.78 $F_{\oplus}$ and 1.91 $F_{\oplus}$ has a high probability of causing a runaway greenhouse [16], resulting in the evaporation of the oceans. However, as already mentioned, we want to be very optimistic from an astrobiological perspective.

Let us specify that habitability is defined by the possible stable presence of liquid water on at least part of the planetary surface [1,2]. In general, it may depend on the ocean's salinity, on the total water mass available, on the sea-level atmospheric pressure, but—most importantly—on the insolation $F$ [1,2].

Once the selection criteria described above are applied, we get a final list of only seven exoplanets: Kepler-62f, Kepler-1126c, Kepler-1544b, Kepler-186f, Kepler-62e, Kepler-442b, and Kepler-440b. The spin evolution of these 7 planets has then been simulated with the *VPLanet* code [8,9].

Before closing this section, it is worth mentioning that the rotation of the host star—unlike that of planets—cannot be subject to a significant slowdown [5], since the star mass is far bigger than the planetary one.

### *2.2. VPLanet Simulator*

*VPLanet v2.5.36* includes different "Physics modules" which can be used to perform different types of astrophysical simulations. The module regarding tidal interactions is called "*eqtide*". It consists of the integration of a group of differential equations, requiring a set of initial conditions to be given as input. Our goal is the computation of the time $\tau$ necessary to achieve a stable tidal trapping (1:1 or 3:2) for the planet due to the gravity field of its host star, starting from the epoch of planetary formation. The general results and features of *eqtide* are as follows [8,9]:

- $\tau$ decreases significantly as the initial orbital eccentricity $e$ of the planet increases;
- The final equilibrium condition is a 1:1 resonance if the initial eccentricity is lower than 0.23; otherwise, it is a 3:2 resonance;
- The evolution of eccentricity $e$ and semi-major axis $a$ values during the time $\tau$ is negligible if the initial value of the acceleration parameter is $w < w_{Mercury}$;
- If $w < w_{Mercury}$, the acceleration parameter $w$ is treated as constant; therefore, one has $\tau \propto w^{-2} \propto T^4_{orbital}$;
- The tidal trapping time is $\tau \propto f_0$, where $f_0$ is the initial spin frequency of the considered planet; therefore, the faster the initial spin, the larger the time needed to achieve tidal trapping;
- $\tau \propto \frac{Q}{k_2}$, where $Q$, $k_2$ are the tidal quality factor of the planet and the second degree Love number of the planet, respectively (see Appendix A);
- The dependence of $\tau$ on planetary mass $M_P$ and planetary radius $R_P$ is weak when compared to its dependence on $w$, $Q$, $k_2$, but it is not negligible;

- The dependence of $\tau$ on stellar specific features (except for $M_{star}$, which determines $w$) is extremely weak when compared to its dependence on $w$, $Q$, $k_2$;
- $Q$, $k_2$, $M_P$, $R_P$, $M_{star}$, and star radius $R_{star}$ are fixed, and their value is constant during the simulation;
- $\tau$ does not depend on the initial planetary obliquity, which is the inclination of the spin axis with respect to the direction perpendicular to the orbital plane.

All these features of the *VPLanet* model are justified in detail in [8,9].

### 2.3. Climatological Considerations

We will compute the expected temperature of planets under the following assumptions:

- Earth-like atmosphere (meaning that the considered atmosphere is identical to that of the Earth, both qualitatively and quantitatively);
- A 2500 m deep global water ocean.

Here, we will refer to these assumptions as "Earth-like" hypotheses, hereinafter abbreviated as "*EL*". In order to reduce the computation time, we will assume in the climatological simulations that the obliquity and the eccentricity are null, even if these are not Earth-like features. Climatological simulations with non-zero obliquity and non-zero eccentricity are reserved for future investigation focusing on the formation of seasonal ice. The choice of an Earth-like initial condition for the atmosphere and the oceans is due to the fact that simulators—including *ExoPlaSim v3.4.0* [17], the code we used—are based on the current Earth to which they are calibrated. Other choices would decrease the reliability of the results. Furthermore, the Earth has a "moderate" greenhouse, more than Mars but less than Venus; hence, the *EL* hypothesis seems to be quite natural. We want to point out that hypothesizing an *EL* initial condition does not mean hypothesizing that the studied planet is inhabited. We are looking for habitability, not hypothesizing it. It is true that some of the features of current Earth atmosphere—like the abundances of $O_2$, $O_3$—are the result of biological processes [18]. But these biologically induced features do not significantly affect habitability. They probably do not affect the climate at all [18]. In fact, most of the climatological models—including the one implemented inside *ExoPlaSim* [17]—consider that greenhousing is provided by carbon dioxide, water vapor, and high-altitude water clouds only, and none of these "actors" are significantly affected by life [18]. It is also true that plants can "store" some carbon by removing $CO_2$ from the atmosphere, but their role is quantitatively devoid of significance, because the real leading actors of the carbon cycle are the oceans, the atmosphere, the land surface, and the planetary interior (which interfaces with the atmosphere through volcanism) [18,19].

We do not expect that our results will give real planetary temperatures, since we do not know the true features of the atmospheres of these planets. However, we believe that our results can be insightful in the context of habitability.

Liquid water under the *EL* hypothesis can exist in the temperature range 273 K–373 K (i.e., 0 °C–100 °C) [6].

The global time-averaged surface temperature of a fast rotator is given by [1,2]:

$$T^4_{fast} = g \cdot (1 - A) \cdot F / (4\sigma) \quad (1)$$

where $g$, $A$, $F$, $\sigma$ are the "greenhouse factor", the global Bond Albedo, the stellar flux arriving at the top of the atmosphere (insolation), and the Stefan–Boltzmann constant, respectively.

We specify that the parameter $g$ only depends on the atmosphere, because it contains all the greenhousing elements (gases and high clouds) [1,2].

Albedo is affected mostly by the oceans (based on ice fraction) but also by the atmosphere (based on low-cloud abundance) [1,2].

Equation (1) can also describe the climate of a 3:2 locked planet, which is a very slow rotator but still "sees" its star moving in the sky. However, for the same insolation $F$ and the same atmospherical–oceanical input (that of the *EL* hypothesis), parameters $g$ and $A$ are expected to evolve differently for a slow rotator and a fast rotator (see Appendix B). Hence, for the slow rotator, we introduce a distinct temperature which we call $T_{slow}$.

Obviously, for a very slow rotator, the choice to define a global time-averaged temperature ($T_{slow}$) is a bit "forced", while it is natural to define $T_{fast}$ for fast rotators. However, the climatological behavior of a 3:2 locked planet is more similar to that of a fast rotator than a 1:1 locked planet, because the position of the host star in the sky—as seen from any physical point on the planetary surface—is not fixed.

The surface temperature of the substellar point on a 1:1 locked planet is given by [1,2]:

$$T_{sub}^4 = \frac{g}{\sigma}(1 - A_{ill}) \cdot F \cdot (1 - 0.75q) \tag{2}$$

where $A_{ill}$ is the Bond Albedo of the illuminated hemisphere, and $q$ is the "heat distribution" factor (which can vary from 0, in case of an absence of atmosphere/oceans, to 1, in case of very dynamical atmosphere/oceans). The surface temperature of the antistellar point on a 1:1 locked planet is given by [1,2]:

$$T_{ant}^4 = \frac{g}{4\sigma}(1 - A_{dark}) \cdot F \cdot q \tag{3}$$

where $A_{dark}$ is the Bond Albedo of the dark hemisphere.

The surface temperature in the dark hemisphere is expected to be almost uniform [1,2].

On the contrary, the temperature in the illuminated hemisphere of a 1:1 locked planet strongly depends on the great circle distance $\theta$ from the substellar point: it has its maximum at $T_{sub} = T_{(\theta=0^\circ)}$ and decreases with increasing $\theta$; the rapidity of the temperature drop with $\theta$ depends on the specific $q$ value [1,2]. Let us clarify that we simply consider the great circle distance $\theta$ from the substellar point without distinguishing between latitude and longitude coordinates, since we are now discussing a 1:1 locked planet with null obliquity. Let us also specify that a 1:1 locked planet is expected to truly have a $\sim 0^\circ$ obliquity [20].

For the selected planets, in the case of 1:1 locking, we will not show $T_{sub}$, $T_{ant}$ but $T_{ill}$, $T_{dark}$ instead, indicating the average temperatures of illuminated and darkened hemispheres, respectively.

We estimate the climate of the selected exoplanets with the software *ExoPlaSim v3.4.0* [17] (see Appendix B for details).

If the planet turns out to not be tidally trapped, it is expected to be a fast rotator, and $T_{fast}$ well describes the planetary climate.

On the other hand, if the planet is tidally trapped, two conditions are possible: 1:1 locking, meaning that two hemispheres have to be distinguished (the illuminated and the darkened ones), and 3:2 locking, meaning that the planet is a very slow rotator but still "sees" the star moving across its sky. In the former case, climate is described by $T_{ill}$ and $T_{dark}$; in the latter case, climate is described by $T_{slow}$.

Unfortunately, due to the absence of information on eccentricity, when a tidal trapping condition for a certain exoplanet is corroborated by our simulations, both $T_{slow}$ and the couple $T_{ill}$, $T_{dark}$ have to be computed, because we cannot know the kind of tidal trapping that has been achieved (3:2 or 1:1). Conversely, if tidal trapping is excluded by our simulations, only $T_{fast}$ has to be computed. In cases where it cannot be determined whether tidal trapping has been achieved, $T_{fast}$, $T_{dark}$, $T_{ill}$, $T_{slow}$ must all be computed. *ExoPlaSim* simulations allow us to compute $T_{fast}$, $T_{dark}$, $T_{ill}$, $T_{slow}$ for a certain planet after the necessary input conditions have been provided to the simulator (see Appendix B).

*2.4. Input Data*

In this section, we show the input parameters, which were provided to *VPLanet* and/or *ExoPlaSim* simulators, for all seven selected exoplanets (see Table 1). We point out that we do not have measurements of the planetary mass $M_P$, but we can estimate it through the mass–radius relation formulated by Tobie et al. [21], because all seven planets have been discovered by the transit method; hence, we have reliable estimates for the radius $R_P$. After the mass has been estimated, the Love number $k_2$ can be evaluated thanks to the mass—$k_2$ relation formulated by Tobie et al. [21]. We also point out that the insolation values $F$ in Table 1 are computed considering the stellar luminosities $L_*$ measured by Stassun et al. [22]. We chose Stassun's catalog as a reference because it is the only one that provides the luminosity values of all the host stars of "our" seven selected exoplanets. The semi-major axis values $a$, which were used to compute the insolation $F$ for each planet—since $F = L_*/(4\pi a^2)$—can be immediately calculated from the provided values of stellar masses and orbital periods through Kepler's third law.

**Table 1.** Selected planets whose spin evolution is simulated. Expected mass and expected $k_2$ values are computed through the mass–radius relation and the $k_2$—mass relation, respectively (see Appendix A), both given by Tobie et al. [21]. Measured values instead of computed values of the mass and of the Love number are only available for Earth and Mercury [5,23]. Insolation values $F$ at distance $a$ from the star are computed from luminosity measurements given by Stassun et al. [22]. Yellow cells refer to data (taken from the literature), while gray cells refer to computed quantities.

| Kepler Name | Radius ($R_\oplus$) | Expected or Real Mass ($M_\oplus$) | Expected or Real ($k_2$) | Star Mass ($M_\odot$) | Orbital Period (Days) | Star Age ($10^9$ Years) | $w$ | $F$ ($F_\oplus$) |
|---|---|---|---|---|---|---|---|---|
| Earth | 1.00 | 1.00 | 0.28 [5,23] | 1.00 | 365.25 | $4.54^{+0.01}_{-0.01}$ [6] | 1.0 | 1.0 |
| 62f | 1.46 [24] | 3.9 | 0.38 | 0.72 [24] | 267 [24] | $2.34^{+2.15}_{-1.02}$ [24] | 2.0 | 0.52 |
| 1126c | 1.68 [25] | 6.4 | 0.41 | 0.92 [24] | 200 [25] | $5.50^{+3.44}_{-3.39}$ [24] | 3.4 | 1.85 |
| 1544b | 1.79 [26] | 8.0 | 0.42 | 0.74 [26] | 169 [26] | $3.90^{+7.30}_{-0.80}$ [26] | 4.7 | 0.92 |
| 186f | 1.12 [24] | 1.5 | 0.31 | 0.51 [24] | 130 [24] | $3.89^{+5.57}_{-2.23}$ [24] | 8.0 | 0.39 |
| 62e | 1.75 [24] | 7.4 | 0.41 | 0.72 [24] | 122 [24] | $2.34^{+2.15}_{-1.02}$ [24] | 9.0 | 1.42 |
| 442b | 1.59 [24] | 5.2 | 0.39 | 0.70 [24] | 112 [24] | $3.09^{+2.72}_{-1.29}$ [24] | 10.9 | 0.92 |
| 440b | 1.63 [24] | 5.7 | 0.40 | 0.55 [24] | 101 [24] | $4.17^{+5.79}_{-2.44}$ [24] | 12.8 | 0.74 |
| Mercury | 0.38 [6] | 0.055 [6] | 0.08 [5,23] | 1.00 | 87.97 [6] | $4.54^{+0.01}_{-0.01}$ [6] | 17.4 | 6.7 |

The energy dissipation parameter $Q$ (also called "tidal quality factor") is expected to have a value of $\sim 10^2$ in the case of rocky planets [8,9,27]. We will keep this value fixed in our simulations, since we assume that it does not change during the time period from $t = 0$ (planetary formation) to $t = \tau$ and not even afterwards [8,9,27].

We note that information about the Love number and tidal quality factor of Earth and other solar system objects comes from geological observations (mapping of gravity and magnetic fields, etc.) [5,21,23].

Simulations of planetary spin evolution have been performed (for each planet) for a set of specific eccentricity values $e$. In particular, we explored the following values: 0.00, 0.08, 0.16, 0.24, 0.32, 0.40. The range 0.00–0.40 is in fact realistic for planets orbiting main sequence stars in our range of $w$ values [28]. We explore this wide range of values because we do not have estimates on the real eccentricity of the selected planets [24–26].

For all *VPLanet* simulations and for all 7 planets, we adopted the following simplifications:

- The initial obliquity of the planet has been set to null value because it is irrelevant to the computation of $\tau$, considering our range of $w$ values [5,8,9];
- The rotation period of the stars has been set to 27 days like the current solar period [5] (however, this parameter is irrelevant to the computation of $\tau$, considering our range of $w$ values [5,8,9]);
- The initial rotation period of the planets has been set to 8 h ($f_0 \sim 3 * 10^{-5}$ Hz), which is the expected value for "newborn" rocky planets [29].

## 3. Results

The results of our simulations are summarized in Table 2.

**Table 2.** Results of spin evolution and climatological simulations. $\tau$ is expressed in units of Gyr (i.e., $10^9$ years). Values of $\tau$ corresponding to $e < 0.23$ are shown in black, while $\tau$ values corresponding to $e > 0.23$ are shown in green. We note that $e = 0.23$ is the threshold separating the 1:1 and the 3:2 resonances as achievable tidal trapping conditions. Simulations regarding Earth (neglecting the Moon) and Mercury have also been performed. $T_{dark}$, $T_{ill}$, $T_{slow}$ are not computed for planets whose simulations yield a fast-rotation condition. $T_{fast}$ is not calculated for planets whose simulations yield a tidal trapping condition. The meaning of cell colors is explained in Section 3. No climatological simulations have been run regarding Mercury, since we know it lacks an atmosphere, so our model would be unsuitable [6]. Please note that the references relative to the star ages (i.e., to the planetary ages) are given in Table 1 and are not reported here.

| Kepler Name, Age (Gyr) | | τ(Gyr) [e=0.00] | τ (Gyr) [e=0.08] | τ (Gyr) [e=0.16] | τ (Gyr) [e=0.24] | τ (Gyr) [e=0.32] | τ (Gyr) [e=0.40] | $T_{fast}$ (K) | $T_{dark}$ (K) | $T_{ill}$ (K) | $T_{slow}$ (K) |
|---|---|---|---|---|---|---|---|---|---|---|---|
| Earth, | 4.54 | ~387 | ~369 | ~326 | ~271 | ~220 | ~176 | ~288 | / | / | / |
| 62f, | 2.34 | ~94.5 | ~90.2 | ~79.3 | ~65.9 | ~53.4 | ~42.9 | ~182 | / | / | / |
| 1126c, | 5.50 | ~34.4 | ~32.8 | ~28.8 | ~24.0 | ~19.4 | ~15.6 | ~357 | / | / | / |
| 1544b, | 3.90 | ~18.2 | ~17.4 | ~15.3 | ~12.7 | ~10.3 | ~8.27 | ~281 | / | / | / |
| 186f, | 3.89 | ~6.15 | ~5.87 | ~5.17 | ~4.29 | ~3.48 | ~2.79 | ~174 | ~120 | ~187 | ~163 |
| 62e, | *2.34* | ~4.83 | ~4.61 | ~4.06 | ~3.37 | ~2.73 | ~2.19 | ~344 | ~320 | ~335 | ~331 |
| 442b, | *3.09* | ~3.26 | ~2.96 | ~2.74 | ~2.27 | ~1.84 | ~1.48 | ~281 | ~234 | ~276 | ~216 |
| 440b, | 4.17 | ~2.35 | ~2.19 | ~1.97 | ~1.64 | ~1.33 | ~1.07 | / | ~173 | ~234 | ~194 |
| Mercury, | 4.54 | ~2.94 | ~2.81 | ~2.47 | ~2.05 | ~1.66 | ~1.33 | / | / | / | / |

In Table 2, blue cells correspond to a fast-rotation condition, since the estimated $\tau$ is greater than the stellar age. Conversely, pink cells correspond to a tidal trapping condition in a 1:1 or 3:2 resonance, since the estimated $\tau$ is smaller than the stellar age.

Note that the estimated stellar ages (see Table 1) have large uncertainties, so if new estimates were to be published in the future, some of our results may need reinterpretation.

Trapping is assumed to be achieved only if the estimated value of $\tau$ is smaller than the planetary age (i.e., the star age). Note that, in *VPLanet* simulations, the rotation period—initially 8 h—first increases at a very slow rate until $\sim 0.95\tau$, after which it rapidly approaches its final value (see Figure 1).

Table 2 is sorted by increasing $w$ and—for each $e$ value—it ends up being sorted by decreasing $\tau$, with a single exception being Mercury, which is very small compared to the other simulated planets. Being so small, it has a significantly lower $k_2$, which explains the peculiarity.

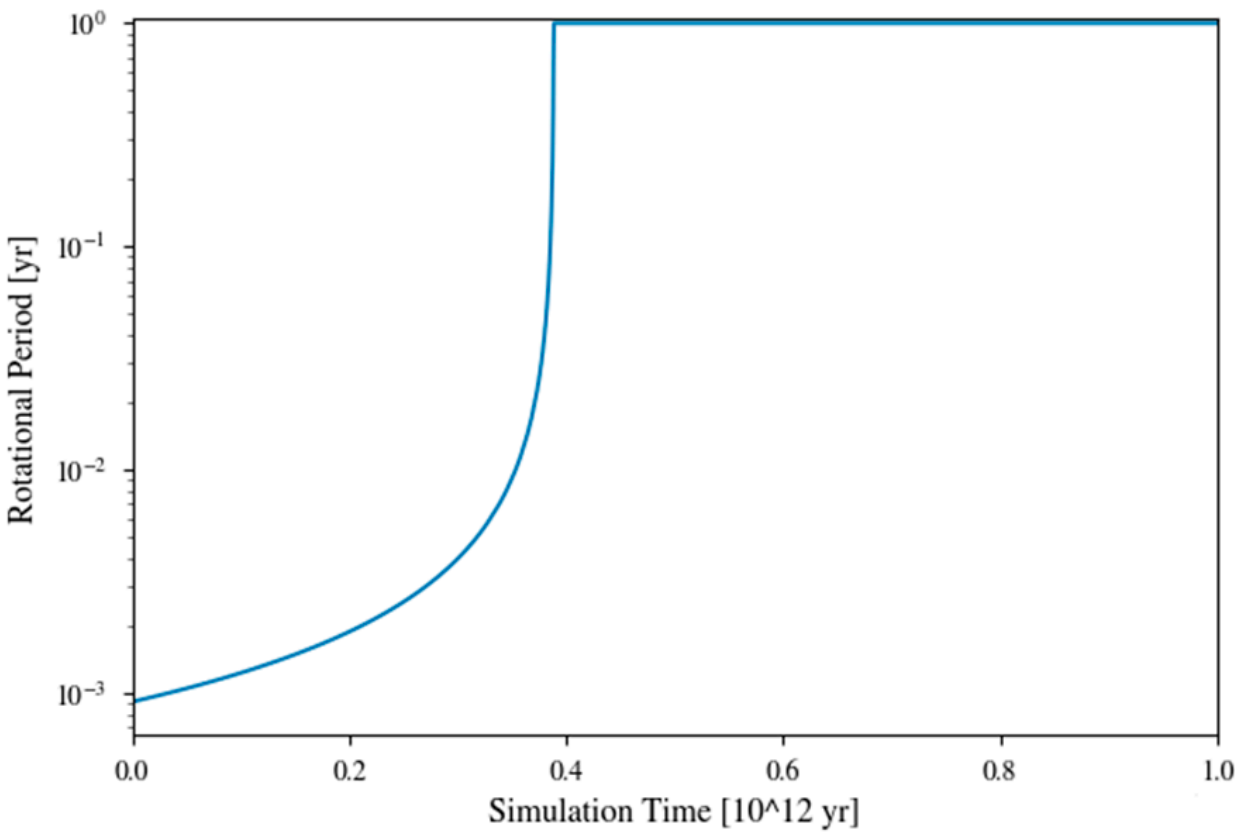

**Figure 1.** Increase in rotational period given by the simulation of Earth's spin evolution in the hypothesis of null eccentricity. Time is expressed in units of $10^{12}$ and $10^0$ Earth years on the horizontal and vertical axes, respectively (1 Earth year = 8766 h). The simulation in this example yields a 1:1 resonance achieved in $\tau = 0.387 \cdot 10^{12}$ yr. The shape of this curve is similar to all the other curves obtained when simulating other planets. The vertical axis is represented with a logarithmic scale, while the horizontal axis is represented with a linear scale.

In Figure 2, we plot the tidal trapping time $\tau$ as a function of the eccentricity for Kepler-62e as an example, exploiting the results presented in Table 2. The shape of this function is practically the same for all the other selected planets, but for brevity, we only show that of Kepler-62e.

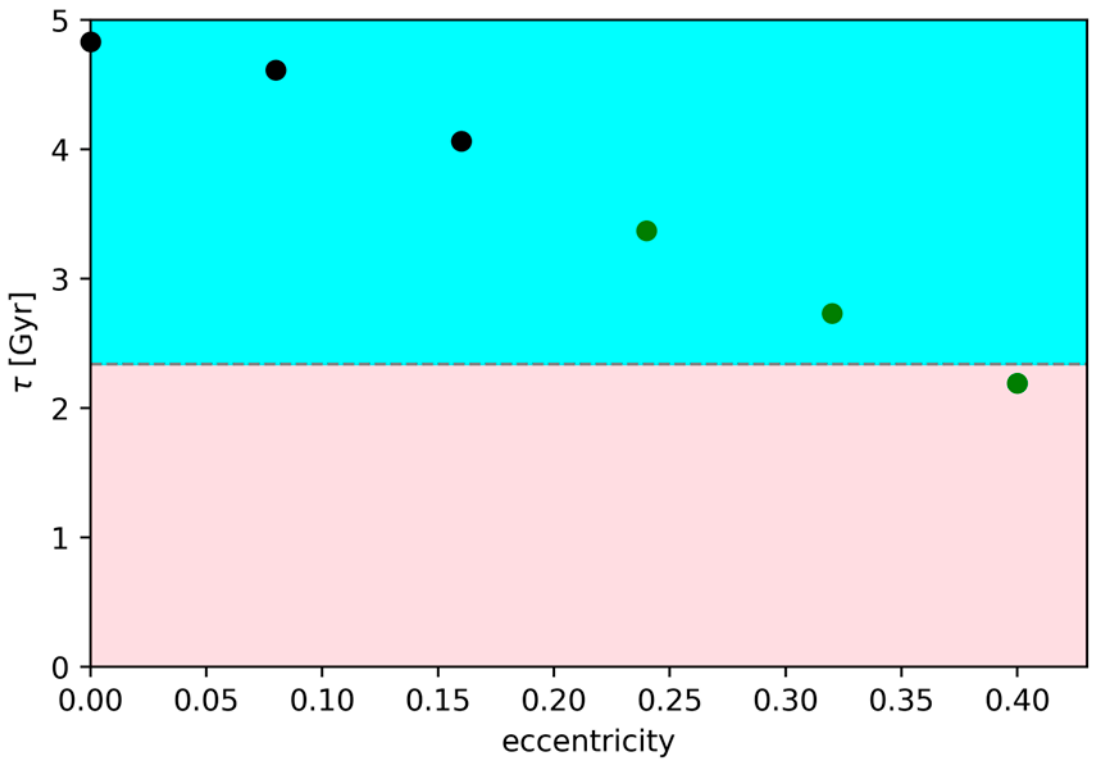

**Figure 2.** Plot of the tidal trapping time $\tau$ as a function of the orbital eccentricity for the planet Kepler-62e. The blue region corresponds to a fast rotation, while the pink region corresponds to a tidal trapping condition. The dashed line dividing the two regions corresponds to the estimated age of the host star.

In Table 2, we see that the simulation results fully agree with the actual spin conditions of Earth and Mercury: Earth is very far from trapping, while Mercury has already achieved a spin–orbit resonance. Mercury was almost certainly born on a more eccentric orbit than the current one: the current eccentricity of ~0.21 [3] would in fact be incompatible with its 3:2 resonance [8,9]. This means that the Sun's tidal forces acting on Mercury have slowed down its rotation and have also slightly circularized its orbit. If a simulation is performed with an initial $e > 0.23$ (the threshold value for 3:2 trapping), a weak but observable circularization is observed for Mercury. Conversely, the decrease in eccentricity (i.e., circularization) is even weaker or totally negligible for lower $w$ values (like those of our seven planets), considering the 0.00–0.40 range of initial eccentricity values.

The temperatures given in Table 2 come from *ExoPlaSim* simulations. Furthermore, it is worth noting that:

- The insolation provided as an input to *ExoPlaSim* is taken from Table 1 for each planet;
- The rotation period provided as input to *ExoPlaSim* in order to compute $T_{fast}$ is always 8 h, which is the expected rotation period of "newborn" rocky planets [29];
- The rotation period provided as input to *ExoPlaSim* in order to compute $T_{slow}$ is simply 2/3 of the orbital period (which is taken from Table 1 for each planet), since we defined the "slow rotator" as a planet trapped in a 3:2 spin–orbit resonance.

Note that, due to the absence of information about eccentricity, when a tidal trapping condition for a certain exoplanet is corroborated by our simulations, both $T_{slow}$ and the couple $T_{ill}$, $T_{dark}$ have been computed, because we cannot know the kind of tidal trapping that has been achieved (3:2 or 1:1). Conversely, if tidal trapping is excluded by our simulations, only $T_{fast}$ is meaningful and has to be computed.

In cases where we could not determine whether the planet is tidally trapped, $T_{fast}$, $T_{dark}$, $T_{ill}$, $T_{slow}$ have all been calculated.

An important consideration resulting from our work is that $T_{slow} < T_{fast}$ for every given insolation value if the *EL* hypothesis holds true. Cloud accumulation and Albedo growth are moderate in the case of imperfect resonance but still stronger compared to a fast rotator, and this explains why $T_{slow} < T_{fast}$ holds true in all cases.

Our study also shows that, in certain cases, $T_{ill} < T_{fast}$: this is counterintuitive but can be explained by Albedo growth, since cloud accumulation is very strong in the case of perfect resonance. In the case of a planet receiving little radiation and getting frozen, cloud accumulation does not happen; hence, $T_{ill} > T_{fast}$, as expected.

## 4. Summary

We have selected seven objects and have tried to resolve the uncertainty about their spin condition. In particular, using the simulator *VPLanet*, we have computed the time $\tau$ needed to achieve tidal trapping, ultimately comparing this value with the host star age. After the current spin condition of a planet has been studied, hypotheses about its climate can be made.

In order to estimate the average temperature of an exoplanet—referring to the temperature of the whole surface or of a single hemisphere, depending on the spin condition—it is not sufficient to know its insolation: it is also necessary to make hypotheses about the features of its atmosphere and oceans. This work takes into consideration the classical *EL* assumption (see Section 2.3), allowing us to perform reliable simulations, even if reducing the generalizability of the results. In fact, *ExoPlaSim* simulations having initial atmospherical conditions strongly different from the current atmosphere of Earth would be less reliable (in particular, when the carbon dioxide concentration is greater than that of water vapor) [17].

Looking at Table 2, we can make some statements about the seven exoplanets. Please note that, as already mentioned, all the climatological results shown in this study are obtained after assuming an initial *EL* (i.e., Earth-like planet) condition.

- Kepler-62f is almost certainly a fast rotator, having a frozen surface; it could be habitable only if it had a far denser atmosphere than what we assumed to prevent freezing.
- Kepler-1126c is almost certainly a fast rotator, with a global average temperature of 357 K (84 °C). Therefore, it appears to be habitable, but we cannot exclude a runaway greenhouse, since we would need to simulate at least ~1 million years of atmospheric evolution [16], which is unfeasible.
- Kepler-1544b is almost certainly a fast rotator, with a global average temperature of 281 K (8 °C), thus appearing as a "cool" but still habitable planet.
- The spin conditions of Kepler-186f, Kepler-62e, and Kepler-442b cannot be determined, since we would need strong constraints on orbital eccentricity, which are currently missing.

We can infer that Kepler-186f cannot sustain liquid water anywhere on its surface, regardless of the spin condition. In fact, a thicker atmosphere would be necessary in order to enhance greenhousing and to prevent the freezing of all the water in the oceans and in the air.

Kepler-62e is expected to be a very warm but habitable planet regardless of the specific spin condition, which, in this case, does not strongly affect the surface temperature.

Kepler-442b is worth a particularly detailed discussion.

If it is a fast rotator, it turns out to be a habitable "cool" planet with an average surface temperature of 281 K = 8 °C.

In the case of 1:1 locking, Kepler-442b should be habitable only in the illuminated hemisphere, and more precisely, near the substellar point.

Conversely, in the case of 3:2 locking, we expect this planet to be entirely frozen, unable to stably sustain liquid water on any point of its surface.

An interesting and apparently surprising outcome of our simulations regarding this planet is that $T_{slow} < T_{dark}$. In the case of the 3:2 spin condition—meaning a very slow rotation—for the adopted value of insolation 0.92 $F_{\oplus}$, a thick ice mantle forms on the night side when the simulation starts. Then, when this mantle gets irradiated by the star thanks to planetary rotation, the ice melting process is very slow because of the high Albedo. The melting is so slow that there is not enough time to defrost the oceans, because planetary rotation will bring the mantle back to the night side before that can happen. In the meantime, another mantle has formed on the other side; hence, the whole planetary surface becomes irreversibly frozen.

On the contrary, the illuminated hemisphere of Kepler-442b in the case of 1:1 locking never becomes totally frozen (for this level of insolation), meaning that the planet has a lower average Albedo, and hence, a higher energy absorption compared to a slow-rotating (i.e., the 3:2-trapped) Kepler-442b. The dark hemisphere of a 1:1 locked Kepler-442b will become frozen, but it will still be warmer than the surface of a 3:2 locked Kepler-442b, since the latter is entirely frozen—resulting in a higher global average Albedo—and receives less net insolation. This difference in energy absorption explains why—for this planet—our simulations indicate that $T_{slow} < T_{dark}$.

- We expect Kepler-440b to be tidally trapped and quite far from being habitable. In fact, climatological simulations yield a totally frozen surface.

**Author Contributions:** Conceptualization: S.C.; Methodology: S.C. and V.O.; Investigation: S.C. and V.O.; Writing—original draft preparation: S.C., V.O., F.D.P. and F.S.; Writing—review and editing: F.D.P., F.S. and L.D.V.; Supervision: F.D.P., F.S. and L.D.V. All authors have read and agreed to the published version of the manuscript.

**Funding:** This research received no external funding.

**Data Availability Statement:** All the data generated by our simulations—both *VPLanet* and *ExoPlaSim* ones—are available upon request (the reader can find the email-addresses on the first page of this paper).

**Acknowledgments:** We want to thank Rory Barnes from the University of Washington who helped us properly run the software *VPLanet v2.5.36*, and Adiv Paradise from the University of Toronto who helped us properly run the software *ExoPlaSim v3.4.0*. F. De Paolis, V. Orofino and F. Strafella acknowledge *project-Euclid (INFN-TAsP)*. S. Camposeo and L. Di Venere acknowledge *INFN (Bari, Italy)* and *Università degli Studi A. Moro (Bari, Italy)* for the use of the high-performance-computing system named *ReCaS* (https://www.recas-bari.it/index.php/en/recas-bari-the-infrastructure, latest access on 26 April 2026).

**Conflicts of Interest:** The authors declare no conflicts of interest.

## Appendix A

The parameter $k_2$, also called the second degree "Love number", describes the mass distribution inside the planet [5,30]: the maximum possible value is 1.5, corresponding to a planet having a uniform density; the minimum possible value is 0, corresponding to a planet having all the mass concentrated in its center. Hence, this parameter—whose complicated mathematical definition will not be discussed here—quantifies how much the mass of a solid body is concentrated near its geometrical center [5,30]. $k_2$ is expected to have a positive correlation with the planetary mass [21].

This correlation can be approximated by the following law—valid for rocky planets—that has been derived from simulations regarding planetary formation [21]:

$$k_2 \sim \frac{M_P^{\frac{2}{3}}}{2M_P^{\frac{2}{3}} + 1.6} \quad \text{(A1)}$$

where planetary mass $M_P$ is expressed in units of Earth mass.

Hence, massive planets are expected to be more uniform than small planets. However, Equation (A1) shows that the theoretical maximum value 1.5 can never really be reached. Let us also point out that the mass–radius relation for rocky planets is estimated to be:

$$R_P \propto M_P^{0.28} \rightarrow M_P \propto R_P^{3.57} \quad \text{(A2)}$$

where both quantities are expressed in Earth units ($R_\oplus$, $M_\oplus$) [21].

The tidal quality factor $Q$ describes the rate of energy dissipation inside the planet due to the tidal interaction between the planetary fluid/rigid matter and the star. In particular [5]:

$$Q = \frac{2\pi E_0}{\Delta E} = \frac{1}{tan(\epsilon)} \quad \text{(A3)}$$

where $E_0$ is the gravitational energy stored inside the tidal bulge (due to its departing from a spherical shape), and $\Delta E$ is the energy dissipated during each orbit; $\epsilon$ (see Figure A1) is the angle between the "bulge line" $\overline{AB}$ and the line connecting the star center to the planet center.

The tidal bulge is simply the deformation of the planetary shape. This deformation arises from the fact that the planet is an extended body, and the star gravity field has slightly different values in A and B (see Figure A1).

In our simulations, we assumed a constant-phase-lag (CPL) model, meaning that $\epsilon$ is assumed to be constant, hence $Q$ is constant too. This is a very usual assumption when tidal interactions are considered [5,8,9]. Another assumption we made is that the orbital motion and the planetary spin are prograde rather than retrograde; it has been demonstrated that this assumption holds true for all planets, except for objects—like Venus—whose rotation has been affected by a giant impact [6,7].

The last assumption is that the initial spin frequency—immediately after the planetary formation—is higher than the orbital frequency [5–7]. Let us note that $\epsilon$ is not null because the spin and the orbital frequencies are unequal; according to the previous assumption that $f_{spin} > f_{orbital}$, the $\overline{AB}$ line constantly "anticipates" the line connecting the two bodies (see Figure A1). Therefore, the gravitational forces—due to the star's gravity field—acting on the A region and on the B region of the bulge are misaligned and produce a torque. While $\epsilon > 0$, the net tidal torque is not null and acts "against" the rotation, causing its slowdown. This slowdown stops when a tidal trapping condition is achieved, because the averaged torque during the orbit is now null [5].

The tidal torque is given by [5]:

$$|\Gamma| = \frac{3Gk_2}{2Q} \cdot w^2 \cdot R_P^5 \tag{A4}$$

where $G$ is Newton's gravity constant, $w$ is the tidal acceleration parameter, and $R_P$ is the planetary radius.

If the spin frequency is greater than the orbital frequency, the torque is negative, leading to the slowdown of the rotation. Otherwise, if the spin frequency is smaller than the orbital one, the torque is positive, leading to the speed up of the planetary spin. If the two frequencies become equal, the torque becomes null; therefore, the rotation period does not change anymore. Let us observe that the torque intensity can be expressed as

$$\Gamma = 2\pi I \cdot \frac{d\left(f_{spin}\right)}{dt} \tag{A5}$$

where $f_{spin}$ is the spin frequency, and $I$ is the polar moment of inertia of the planet.

So, from Equations (A4) and (A5):

$$\left|\frac{d\left(f_{spin}\right)}{dt}\right| = \frac{3Gk_2}{2Q} \cdot w^2 \cdot R_P^5 \cdot \frac{1}{2\pi I} = \frac{3Gk_2 w^2}{4\pi Q} \cdot \frac{R_P^5}{\alpha M_P R_P^2} \tag{A6}$$

where $\alpha$ is a positive number, and $M_P$ is the planetary mass. From Equations (A2) and (A6), we easily obtain that the direct dependence of the slowdown on the planetary radius almost cancels out. A rough estimate of the dependence of $\tau$ on the planetary mass can be given by Equations (A2) and (A6):

$$\tau \sim \frac{f_0}{\left|\frac{d\left(f_{spin}\right)}{dt}\right|} \propto \frac{M_P R_P^2}{R_P^5} \sim R_P^{0.57} \sim M_P^{0.28 \cdot 0.57} \sim M_P^{0.16} \tag{A7}$$

where $f_0$ is the initial spin. Equation (A7) demonstrates that the dependence of $\tau$ on $M_P$ is very weak.

Let us conclude this brief review of tidal physics with an observation about energy conservation. As the slowdown of the rotation corresponds to a decrease in rotational energy, we may ask where this energy has gone to. The answer is that it has been converted into internal heat thanks to the friction between the external bulge and the underlying layers.

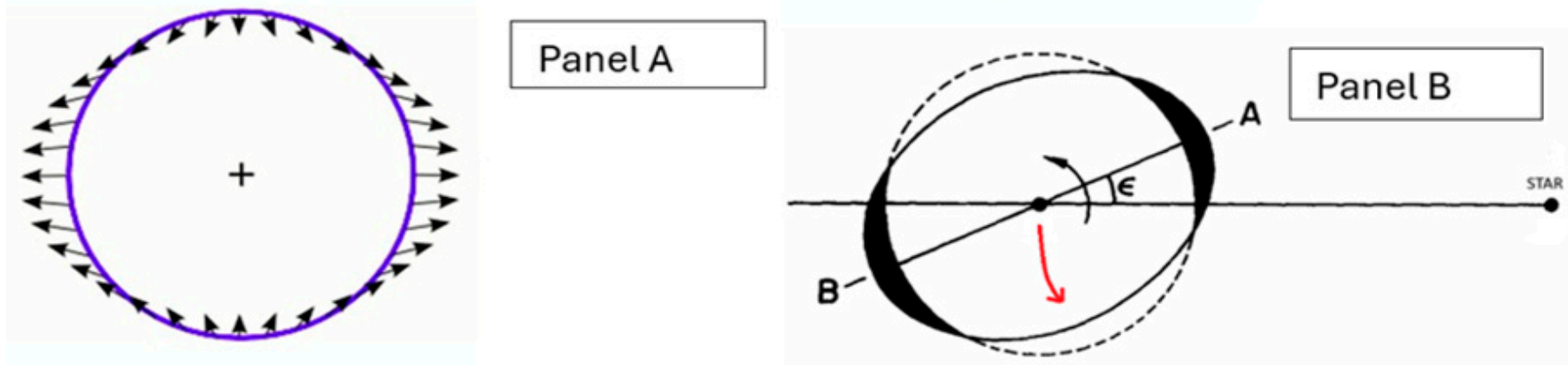

**Figure A1.** PANEL (**A**): Diagram of the net tidal accelerations acting on the planetary surface (caused by the host star's gravity field). The planet is represented with a top view (i.e., as seen from above). Accelerations are computed in the reference frame of the planetary center. PANEL (**B**): Representation of the tidal bulge. The planet is represented with a top view. Planetary spinning (black arrow) is counterclockwise, as is the orbital motion around the star (red arrow).

## Appendix B

*ExoPlaSim v3.4.0* [17] is a Fortran/Python code package which implements a set of climatological "rules" and allows the user to simulate the climate of a planet after the spin condition, insolation, and features of the atmosphere and oceans have been given as input.

In our case, the atmospheric/oceanic input is that of the *EL* hypothesis. Let us observe that the *EL* is a set of initial conditions, but features of the atmosphere and of the oceans can change during the simulation.

For example, a very cold planet will obviously freeze all the water vapor in the atmosphere and all the liquid water in the oceans; this would also lead to an Albedo growth, meaning a further cooling. On the other hand, a very hot planet quickly vaporizes its oceans, leading to an increase in water vapor greenhousing, meaning further heating. Hence, input values of Albedo and $g$ (and eventually $q$) are those of an Earth-like planet, but they are free to change during the simulation. Equations (1)–(3) (see Section 2.3) show the dependence of the surface temperature on many parameters, which are not entirely independent of each other.

*ExoPlaSim* simulations permit us to compute $T_{fast}$, $T_{dark}$, $T_{ill}$, $T_{slow}$ for a certain planet after the needed input conditions have been provided. Fifty-year simulations have been run, because a long time is necessary to achieve the thermal equilibrium [17]. A. Paradise et al. [17] have showed—through a comparation with other climate simulators—that for fast rotators with Earth-like atmoshpere, oceans and insolation, a warm-bias of a few Kelvins is expected from *ExoPlaSim* results. For example, longer simulations (>200 years) would yield for the Earth a slightly greater temperature than that showed in Table 2. Long-duration *ExoPlaSim* simulations are very difficult to be achieved because they tend to crash—particularly when the insolation is greater than 1 $F_{\oplus}$. We defer to a future work the possibility to implement longer climate simulations, hopefully with a new version of *ExoPlaSim*.

It is important to note that *ExoPlaSim* produces a temperature map as its output, in which an orbit-averaged temperature—computed considering only the last of the simulated years—is associated with each spatial coordinate. In particular, 64 different values of longitude and 32 different values of latitude are taken into account, forming a grid of 1953 cells.

The 3:2 trapped planet is climatologically similar (but not equivalent) to the fast rotator, since the only difference is the rotation period provided as input.

All the $T_{fast}$ values in Table 2 are obtained by assuming an 8 h rotation period. But ExoPlaSim simulations show stable results in the wide range of 1 h–100 h of the rotation period values; in fact, the discrepancy is always $< 3$K with respect to the $T_{fast}$ resulting from an 8 h input. Hence, the specific value of 8 h does not constrain the validity of the simulation results.

## References

1. Wandel, A. Habitability and sub glacial liquid water on planets of M-dwarf stars. *Nat. Commun.* **2023**, *14*, 2125. [CrossRef] [PubMed]
2. Wandel, A.; Gale, J. The bio-habitable zone and atmospheric properties for planets of red dwarfs. *Int. J. Astrobiol.* **2020**, *19*, 126–135. [CrossRef]
3. Noyellesa, B.; Frouardb, J.; Makarovc, V.; Efroimsky, M. Spin–orbit evolution of Mercury revisited. *Icarus* **2014**, *241*, 26–44. [CrossRef]
4. Williams, G. Geological Constraints on the Precambrian History of Earth's Rotation and the Moon's Orbit. *Rev. Geophys.* **2000**, *38*, 37–59. [CrossRef]
5. Souchay, J.; Mathis, S.; Tokieda, T. *Tides in Astronomy and Astrophysics*, 1st ed.; Springer: Berlin/Heidelberg, Germany, 2013.
6. Faure, G.; Mensing, T. *Introduction to Planetary Science*, 1st ed.; Springer: Berlin/Heidelberg, Germany, 2007.
7. Raymond, S.; Schlichting, H.; Hersant, F.; Selsis, F. Dynamical and collisional constraints on a stochastic late veneer on the terrestrial planets. *Icarus* **2013**, *226*, 671–681. [CrossRef]
8. Barnes, R.; Luger, R.; Deitrick, R.; Driscoll, P.; Quinn, T.R.; Fleming, D.P.; Smotherman, H.; McDonald, D.V.; Wilhelm, C.; Garcia, R.; et al. VPLanet The Virtual Planet Simulator. *Publ. Astron. Soc. Pac.* **2020**, *132*, 024502. [CrossRef]
9. Barnes, R. Tidal locking of habitable exoplanets. *Celest. Mech. Dyn. Astron.* **2017**, *129*, 509–536. [CrossRef]

10. Clouse, C.; Ferroglia, A.; Fiolhais, M. Spin-orbit gravitational locking—An effective potential approach. *Eur. J. Phys.* **2022**, *43*, 035602. [CrossRef]
11. Nakajima, M.; Genda, H.; Asphaug, E.; Ida, S. Large planets may not form fractionally large moons. *Nat. Commun.* **2022**, *13*, 568. [CrossRef]
12. Batalha, N. Exploring exoplanet populations with NASA's Kepler Mission. *Proc. Natl. Acad. Sci. USA* **2014**, *111*, 12647–12654. [CrossRef]
13. Otegi, J.; Bouchy, F.; Helled, R. Revisited mass-radius relations for exoplanets below 120 $M_{\oplus}$. *Astron. Astrophys.* **2020**, *634*, A43. [CrossRef]
14. Lammer, H.; Selsis, F.; Chassefière, E.; Breuer, D.; Grießmeier, J.-M.; Kulikov, Y.N.; Erkaev, N.V.; Khodachenko, M.L.; Biernat, H.K.; Leblanc, F.; et al. Geophysical and atmospheric evolution of habitable planets. *Astrobiology* **2010**, *10*, 45–68. [CrossRef]
15. Jinno, T.; Saitoh, T.; Ishigaki, Y.; Makino, J. N-body simulation of planetary formation through pebble accretion in a radially structured protoplanetary disk. *Publ. Astron. Soc. Jpn.* **2023**, *75*, 951–969. [CrossRef]
16. Kasting, J.; Kopparapu, R.; Ramirez, R.; Harman, C. Remote life-detection criteria, habitable zone boundaries, and the frequency of Earth-like planets around M and late K stars. *Proc. Natl. Acad. Sci. USA* **2014**, *111*, 12641–12646. [CrossRef] [PubMed]
17. Paradise, A.; Macdonald, E.; Menou, K.; Lee, C.; Fan, B.L. ExoPlaSim: Extending the Planet Simulator for exoplanets. *Mon. Not. R. Astron. Soc.* **2022**, *511*, 3272–3303. [CrossRef]
18. Walker, J.C. How life affects the atmosphere. *BioScience* **1984**, *34*, 486–491. [CrossRef]
19. Sleep, N.; Zahnle, K. Carbon dioxide cycling and implications for climate on ancient Earth. *JGR Planets* **2001**, *106*, 1373–1399. [CrossRef]
20. Heller, R.; Leconte, J.; Barnes, R. Tidal obliquity evolution of potentially habitable planets. *Astron. Astrophys.* **2011**, *528*, A27. [CrossRef]
21. Tobie, G.; Grasset, O.; Dumoulin, C.; Mocquet, A. Tidal response of rocky and ice-rich exoplanets. *Astron. Astrophys.* **2019**, *630*, A70. [CrossRef]
22. Stassun, K.G.; Oelkers, R.J.; Paegert, M.; Torres, G.; Pepper, J.; De Lee, N.; Collins, K.; Latham, D.W.; Muirhead, P.S.; Chittidi, J.; et al. The Revised TESS Input Catalog and Candidate Target List. *Astron. J.* **2019**, *158*, 138. [CrossRef]
23. Breuer, D.; Spohn, T.; Van Hoolst, T.; van Westrenen, W.; Stanley, S.; Rambaux, N. Interiors of Earth-Like Planets and Satellites of the Solar System. *Surv. Geophys.* **2022**, *43*, 177–226. [CrossRef]
24. Morton, T.D.; Bryson, S.T.; Coughlin, J.L.; Rowe, J.F.; Ravichandran, G.; Petigura, E.A.; Haas, M.R.; Batalha, N.M. False positive probabilities for all kepler objects of interest: 1284 newly validated planets and 428 likely false positives. *Astrophys. J.* **2016**, *822*, 86. [CrossRef]
25. Valizadegan, H.; Martinho, M.J.S.; Jenkins, J.M.; Caldwell, D.A.; Twicken, J.D.; Bryson, S.T. Multiplicity Boost of Transit Signal Classifiers: Validation of 69 New Exoplanets Using the Multiplicity Boost of ExoMiner. *Astron. J.* **2023**, *166*, 28. [CrossRef]
26. Torres, G.; Kane, S.R.; Rowe, J.F.; Batalha, N.M.; Henze, C.E.; Ciardi, D.R.; Barclay, T.; Borucki, W.J.; Buchhave, L.A.; Crepp, J.R.; et al. Validation of Small Kepler Transiting Planet Candidates in or near the Habitable Zone. *Astron. J.* **2017**, *154*, 264. [CrossRef]
27. Driscoll, P.; Barnes, R. Tidal Heating of Earth-like Exoplanets around M Stars: Thermal, Magnetic, and Orbital Evolutions. *Astrobiology* **2015**, *15*, 739–760. [CrossRef]
28. Sagear, S.; Ballard, S. The orbital eccentricity distribution of planets orbiting M dwarfs. *Proc. Natl. Acad. Sci. USA* **2023**, *120*, e2217398120. [CrossRef]
29. Harris, A.; Ward, W. Dynamical Constraints on the Formation and Evolution of Planetary Bodies. *Annu. Rev. Earth Planet. Sci.* **1982**, *10*, 61–108. [CrossRef]
30. Kramm, U.; Nettelmann, N.; Redmer, R.; Stevenson, D. On the degeneracy of the tidal Love number k2 in multi-layer planetary models: Application to Saturn and GJ436b. *Astron. Astrophys.* **2011**, *528*, A18. [CrossRef]